\documentclass[12pt,preprint2]{aastex}

\newcommand{\cpd}{CPD--56$^{\rm o}$8032}

\slugcomment{To appear in ApJ Letters}

\shorttitle{A dust disk around \cpd }
\shortauthors{De Marco et al.}

\begin{document}

\title{Discovery of an edge-on dust disk around the [WC10] central star 
CPD--56$^{\rm o}$8032}

\author{Orsola De Marco\altaffilmark{1}, M. J. Barlow\altaffilmark{2} and
M. Cohen\altaffilmark{3}}
\altaffiltext{1}{Dept. of Astrophysics, American Museum of Natural History,
   New York, NY 10024}
\altaffiltext{2}{Dept. of Physics and Astronomy, University College 
London, London WC1E 6BT, UK}
\altaffiltext{3}{Radio Astronomy Laboratory, University of California, 
Berkeley, CA 94720}

\begin{abstract}
We present Hubble Space Telescope ultraviolet and optical STIS 
spectroscopy of
the [WCL] planetary nebula central star \cpd ,
obtained during its latest lightcurve minimum. The UV spectrum shows the
central star's continuum light distribution to be split into two bright
peaks separated by 0.10~arcsec. We interpret this finding as due to an
edge-on disk or torus structure that obscures direct light from
the star, which is seen primarily via its light scattered from the disk's rims or lobes.
\cpd\ is an archetype of dual dust chemistry [WCL] planetary nebulae,
which exhibit strong infrared emission features from both carbon-rich and
oxygen-rich materials, and for which the presence of a disk harboring the
O-rich grains had been suggested. Our direct observation of an edge-on
occulting dust structure around \cpd\ provides strong support for such a 
model and for binary interactions being responsible for the correlation
between the dual dust chemistry phenomenon in planetary nebulae and the
presence of a hydrogen-deficient [WCL] Wolf-Rayet central star.
 
\end{abstract}


\section{Introduction}

WC Wolf-Rayet central stars of planetary nebulae ([WC] CSs of PNe) are
H-deficient stars that exhibit strong ionic emission lines of helium,
carbon and oxygen from their dense stellar winds. Amongst the coolest
central stars in this group are \cpd\ (the nucleus of the PN He~3-1333)
and the CS of He~2-113 (both classified as [WC10], Crowther et al. 1998).
Cohen et al. (1986) found the mid-infrared KAO spectra of both these
objects to show very strong unidentified infrared bands (UIBs -- usually
attributed to polycyclic aromatic hydrocarbons, PAHs). Indeed, both the
nebular C/O ratios (De Marco, Barlow \& Storey 1997) and the ratio of UIB
luminosity to total IR dust luminosity (Cohen et al. 1989) for these two
objects are among the largest known. It was therefore a major surprise
when mid- and far-IR Infrared Space Observatory spectra of these two
objects showed the presence of many emission features longwards of
20~$\mu$m that could be attributed to crystalline silicate and water ice
particles (Barlow 1997, Waters et al.  1998a, Cohen et al. 1999),
indicating a dual dust chemistry, i.e.  the simultaneous presence of both
C-rich dust and O-rich dust. The dual dust chemistry phenomenon in PNe
appears to show a strong correlation with the presence of a late WC
([WCL]) nucleus -- four out of six [WC8-11] nuclei studied by Cohen et al.
(2002) showed similar dual dust chemistries. In the context of a single star
scenario, this would point to a recent transition (within the last
$\sim$1000~yr) between the O- and the C-rich surface chemistries. However,
the probability of finding a post-AGB object that had recently changed
from an O-rich to a C-rich surface chemistry due to a third dredge-up
event should be very low indeed. An alternative scenario envisages these
systems as binaries (Waters et al. 1998a, Cohen et al. 1999, 2002), in
which the O-rich silicates are trapped in a disk as a result of a past
mass transfer event, with the C-rich particles being more widely
distributed in the nebula as a result of recent ejections of C-rich
material by the nucleus. Because of the quasi-periodic light variations
shown by \cpd , Cohen et al. (2002) suspected the presence of a precessing
disk around it.  In this Letter we present the first direct evidence for
an edge-on disk or torus around \cpd, as revealed by recent HST STIS
spectroscopy.

\section{Observations}

\begin{figure*}
\vspace{5cm}
\includegraphics{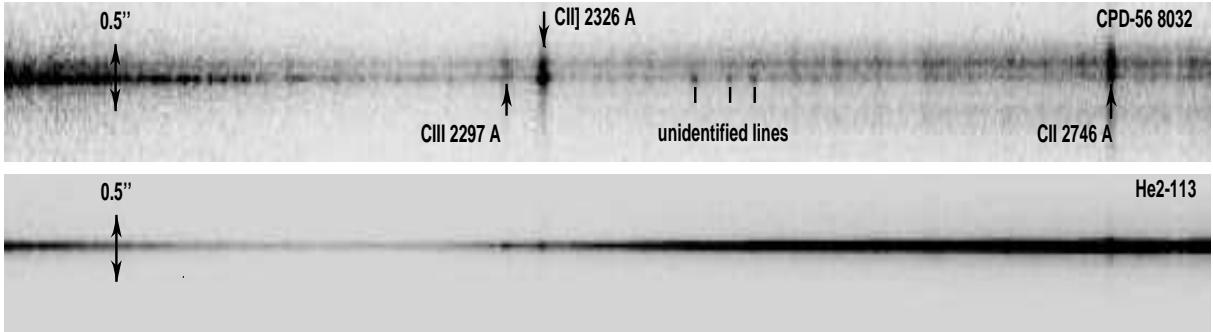}
\caption{STIS/MAMA 2-D spectra of the planetary nebulae He~3-1333 (top;
central star = \cpd\ = V837~Ara) and He~2-113 (bottom;
central star = Hen~1044).}
\label{fig:cpd_hen_ds9}
\end{figure*}

\begin{figure}
\vspace{5cm}
\includegraphics{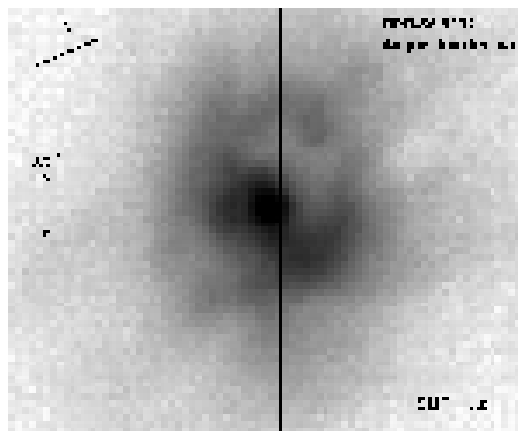}
\includegraphics{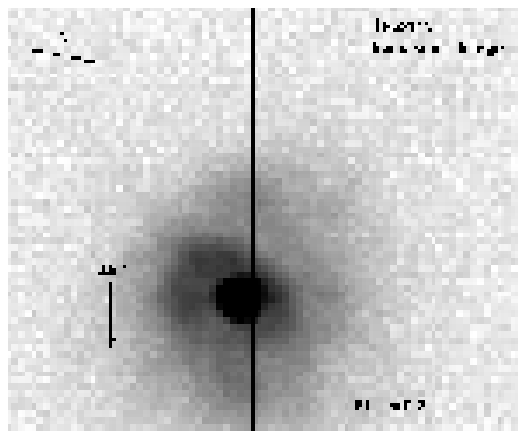}
\caption{STIS long-pass filter ($\lambda >
5500$~\AA ) acquisition images (exposure time 5.8~sec for CPD-56$^{\rm
o}$8032, top; 3~sec for He~2-113, bottom).}
\label{fig:cpd_hen_ds9_2}
\end{figure}

STIS/HST observations of \cpd\ (and of He~2-113 and SwSt~1; De Marco et
al. in preparation) were obtained in 2001 April 1-5, when the star was in
its third recorded visual light minimum (V$\sim$12.5~mag;  Cohen et al.  
2002).  The MAMA+G230L grating (R$\sim$740, $\lambda$=1570-3180~\AA ,
exposure time 2200~sec); CCD+G430M grating (R$\sim$6600,
$\lambda$=3540-3820~\AA , exposure time 1120~sec) and CCD+G750L grating
(R$\sim$800, $\lambda$=5240-10\,270~\AA , exposure time 910~sec) were used
together with the 52$\times$0.2 arcsec$^2$ aperture. In the top panel
of Fig.~\ref{fig:cpd_hen_ds9} we present the 2-dimensional MAMA spectral
image of \cpd\ (0.024 arcsec/pix in the spatial direction). 
For comparison, we also present similar observations made of
the [WC10] nucleus of He~2-113 and its nebula (lower panel). In 
Fig.~\ref{fig:cpd_hen_ds9_2} we present 
the long-pass filter ($>$5500~\AA ) CCD acquisition images (F28X50LP; 0.050
arcsec/pix) for both objects. The slit was
oriented with a position angle (PA) of 243~deg for \cpd\ and
300~deg for He~2-113. The near-UV spectral image of He~2-113 in Fig.~1 shows a spatially
unresolved continuum attributable to the central star, while that of \cpd\
shows the continuum light distribution to be split into two bright peaks
separated by 0.10~arcsec, together with a third, weaker, peak displaced by
0.3~arcsec to the northeast. In the much longer wavelength direct image of
\cpd , such structure cannot be resolved.
The morphology of \cpd\ (Fig.~\ref{fig:cpd_hen_ds9_2})
is highly uncertain
(the only previous HST images were pre-COSTAR; De Marco et al. 1997).  
If any symmetry axis is present, it lies at about PA$\sim$10~deg, in
which case the slit PA would lie at about 50~deg to the symmetry
axis. Our long-pass filter acquisition image appears to be at too long a 
wavelength (central wavelength $\sim$7200~\AA ) for any preferred axis to 
be discernible in the extended profile of the central object.

\section{Results}

\begin{figure}
\vspace{6cm}
\includegraphics{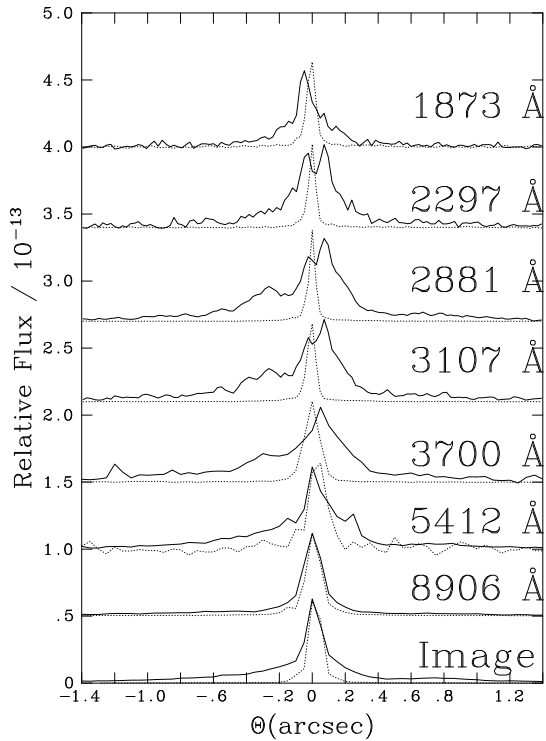}
\caption{STIS/MAMA and CCD `cross-cuts' in the spatial direction, comparing
\cpd 's intensity distribution (solid) to that of He~2-113 at the
same wavelength (dotted, scaled to \cpd 's peak intensity). The wavelength
of each cut is indicated. \cpd 's cuts are normalized to the peak
intensity at 2297~\AA\ and the zero levels are successively incremented by
0.5$\times$10$^{-13}$ ergs~cm$^{-2}$~s$^{-1}$~\AA $^{-1}$.
The bottom curve is a cut at PA = 243~deg through the acquisition
image ($\lambda >$ 5500~\AA , central wavelength 7200~\AA ).}
\label{fig:cuts}
\end{figure}

In Fig.~\ref{fig:cuts} we present spatial cross-cuts (where 10 MAMA/CCD columns were
averaged) through the \cpd\ and
He~2-113 spectra, centered at different wavelengths.
The spatial profile
through He~2-113's central star is very close to that expected for the
MAMA+G140L and CCD+G430M and G750L PSFs (as detailed in the STIS handbook), while
\cpd\ exhibits a much broader profile, which in the MAMA spectrum is
dominated by two peaks separated by 4 pixels = 0.096~arcsec. The similarity
of the spatial distribution of \cpd 's stellar continuum, 
at 3107~\AA\ (MAMA) and at 3700~\AA\ (CCD), implies 
that the splitting observed in the MAMA
spectrum is not due to an observational peculiarity. The
Fig.~\ref{fig:cuts} MAMA cross-cuts show the northeasterly of the two
sharp peaks to be brighter at the shortest UV wavelengths, with the
southwesterly peak brightening to be stronger at the longest UV
wavelengths. The two main peaks seen in the UV are no longer resolved in
the CCD optical cross-cuts, due to the two times larger dimensions of the
0.05~arcsec CCD pixels and the linear increase with wavelength of the
HST's diffraction-limited angular resolution (0.093~arcsec at 8906~\AA ).
The acquisition image has a filter central wavelength of 7200~\AA\ and the
profile of its PA = 243~deg cross-cut appears intermediate between those
that we made at 5412~\AA\ and 8906~\AA\ in the CCD spectral image. The
\cpd\ spatial profile at 8906~\AA\ is only slightly broader than that for
He~2-113, indicating that \cpd 's central star is maybe being seen directly
at that wavelength, with the reflected light components having decreased
significantly in intensity.

From Fig.~\ref{fig:cuts} it is clear that the reflected light spectrum is
a function of spatial position. In particular, to the northeast, at about
--0.3~arcsec, the reflected light increases in intensity at around
3000~\AA , relative to the two main peaks (see the 2881~\AA\ cross-cut in
Fig.~\ref{fig:cuts}). We also note that the lower of the two principal
reflected continuum peaks exhibits three unidentified emission lines, at
2439, 2466 and 2483~\AA\ (see upper panel of Fig.~\ref{fig:cpd_hen_ds9}),
indicating that an emission component, perhaps fluorescent, must be
present in addition to the stellar scattered-light component. The detailed
spectral properties of the individual regions discernible in our STIS
spectral images will be discussed elsewhere.

A comparison between 1800--3200~\AA\ low resolution IUE spectra taken in
1979/80, July 1987 and August 1991 shows the IUE spectra all to have had
identical spectral shapes, though the absolute flux levels in 1987 and
1991 were 0.72 and 0.80 of those in 1979/80 (the UBV photometry of
Pollacco et al. 1992 shows that \cpd\ was at its maximum brightness level
when the August 1991 IUE spectra were acquired). A summation of all the
main emission regions discernible along the length of the 0.2-arcsec wide
MAMA slit (extracted using a 158-pixel wide aperture between pixel 590 and
679 in the 2D STIS image) yielded a spectrum which longwards of 2200~\AA\
has the same spectral shape as the IUE spectra but which becomes
progressively bluer than the IUE spectra shortwards of 2200~\AA . A change
in stellar parameters is excluded by the fact that the UV and optical
stellar emission lines in the STIS spectra have very similar equivalent
widths to those measured on the IUE and AAT spectra of De Marco et al.
(1997). However, a more detailed comparison between the IUE and STIS
spectra is hindered by the fact that the area of the
20$\times$10~arcsec$^2$ IUE entrance aperture was considerably larger than
that of the 0.2~arcsec-wide STIS long slit used in May 2001, so that the
latter is likely to have sampled only a fraction of the total emission --
longwards of 2200~\AA\ the continuum flux level in the summed MAMA
spectrum is 18 times lower than that in the 1979/80 IUE spectra.

\section{Discussion}

Our working hypothesis is that the observed spatial splitting of \cpd 's
continuum (Fig.~\ref{fig:cpd_hen_ds9} and \ref{fig:cuts})  
is due to obscuration of the central object by a
dusty disk or torus seen close to edge-on. Under this interpretation, the
two continuum peaks separated by 0.10~arcsec are attributed to scattered
light emerging from the upper and lower rims of the disk/torus, or from
two lobes residing above and below the disk/torus, 
resembling
the edge-on structure observed in the Red Rectangle, HD~44179 (Osterbart,
Langer \& Weigelt 1997) or the edge-on silhouette disk Orion 114-426
(McCaughrean et al. 1998). It seems possible that the STIS slit was
oriented nearly normal to the PA of the disk, in agreement with the slit
PA being only $\sim$50~deg from the inferred axis of symmetry of the
nebula. Our May 2001 STIS spectrum was obtained during a light minimum of
\cpd , the latest of three observed visual light declines having an
apparent periodicity of $\sim$5~yr (Cohen et al. 2002). Cohen et al.
suggested that these light declines could be due to the precession of a
nearly edge-on obscuring disk, or alternatively that they could be due to
occultations of the central object by particularly dense clumps orbiting
within the disk, in which case the apparent periodicity might be
accidental.  One might argue that \cpd's light variations could be due to
random dust ejections, similar to those of R Coronae Borealis (R~CrB)
stars. Although \cpd 's apparent 5-year period can be confirmed only by
more observations, we note that R~CrB light declines are much deeper
($\sim$8~mag) and always associated with spectral variability (Clayton
1996). Even V348 Sgr (a [WC]-like CS not dissimilar to \cpd ; Leuenhagen,
Heber \& Jeffery 1994) has deep declines and spectral variability,
similarities to R~CrB characteristics which are not shared by \cpd .
Finally, resolved occulting structures, of the type found here around \cpd
, have never been detected around R~CrB stars.

As an alternative to disk precession being responsible for \cpd 's visible
light variability, the orbital motion of the star itself (clearly we have
to assume the presence of a binary companion) might bring it in and out of
alignment with a denser region in the disk. A similar behavior has been
inferred by Van Winckel et al. (1999) for the binary post-AGB objects
HD~52961 and HR~4049, for both of which circumstellar extinction
variations were found to occur on the same timescales as the respective
binary orbital periods of 1310~days and 429~days, though we note that
their optical extinction variations show amplitudes of only
0.16--0.20~magnitudes, versus 1.6~magnitudes for \cpd . If we assume the
alleged companion star to have a similar mass to \cpd\ (and that both have
masses of 0.6~M$_\odot$), the orbital semi-major axis, for a circular
orbit, would be 3.1~AU. The 0.10~arcsec angular separation between the two
main scattered light components in the MAMA spectral image, are assumed to
correspond to the top and bottom edges of a disk or torus with
a projected disk semi-thickness of 67~AU, (for a distance of 1.35~kpc; De
Marco et al. 1997) or to two lobes residing at 67~AU above and below the disk/torus. 
Cohen et al. (1999) estimated the 65-90~K oxygen-rich
grains around \cpd\ to lie 1000~AU from the star, but this was based on an
optically thin assumption. Our present results indicate that these cool
O-rich grains are likely to be located much closer to the star, in the
outer regions of an optically thick edge-on reprocessing disk. As noted
above, a similar structure to that of \cpd\ is observed around the
HD~44179 nucleus of the Red Rectangle, a non H-deficient post-AGB object 
that was found by Waters et al. (1998b) to exhibit a dual dust chemistry.
Inspection of Figs.~2 and 3 of Cohen et al. (2002) shows the 
long wavelength peak of \cpd 's infrared energy distribution to 
lie at the same wavelength as those of the three other [WCL]
objects plotted, but also that this peak is much weaker relative to
the 7.7~$\mu$m UIB-like feature than is the case for the other
three objects. If the crystalline silicate and long-wavelength 
continuum emission originates from an orbiting disk-like region
around each object then this effect could be due to \cpd 's disk
being viewed edge-on, thereby significantly reducing the amount
of radiation received from its disk.

One remaining issue to be considered is the very strong association
between the dual dust chemistry phenomenon in PNe and the presence of a
H-deficient central star, and what that might tell us about a common cause
for the two phenomena. We might expect that some [WC] stars are
single and simply evolve to be a [WC] CS as a result of the well-timed
thermal pulses described by Herwig (2001), in which case they would not be
expected to exhibit dual dust chemistries. Indeed, several [WCL] PNe show 
no evidence for dual dust chemistries
(Cohen et al. 2002, Hony et al. 2001) and may well have originated via the
well-timed thermal pulse mechanism. However, the current evidence points
to binarity playing a key role in the origin of at least some H-deficient
[WCL] 
nuclei {\em and} the dual dust chemistry phenomenon, with mass transfers
first leading to the creation of a massive circum-binary disk and later
stripping most or all of the H-rich surface layer from the CS progenitor.
Under this scenario, the B9/A0~III central star of the Red Rectangle might
be interpreted as possessing only a very thin surface hydrogen layer,
which will quickly be stripped away to yield a H-deficient [WCL] nucleus
once the stellar effective temperature increases to high enough values
($>$ 20,000~K) for radiation pressure driven mass loss to become
significant.

To conclude, our STIS observations indicate the presence of an edge-on
occulting dust structure in the \cpd\ system -- further observations will
be needed to determine how its appearance changes as the star returns to
maximum. It would also be desirable for the orbital characteristics of the
inferred \cpd\ binary system to be determined directly via radial velocity
measurements of the [WC10] star's emission lines. Similar radial velocity
measurements should also be obtained for the central stars of
other dual dust chemistry [WCL] PNe, even if, as is likely in the case of 
He~2-113, the possible dust disks are not oriented edge-on to us.

\acknowledgments

OD gratefully acknowledges Janet Jeppson Asimov 
for financial support. OD and MC acknowledge support from
NASA grant HST-GO-08711.05-A.
Our thanks go to Paul Goodfroj, STIS scientist,
for helping us to understand the instrumental intricacies of STIS.

\clearpage

\end{document}